\documentclass[aps,pra,twocolumn,showpacs]{revtex4}

\usepackage{graphicx}

\usepackage{bm}

\begin{document}

\title{Distillation and purification of symmetric entangled Gaussian states}

\author{Jarom\'{\i}r Fiur\'{a}\v{s}ek} 
\affiliation{Department of Optics, Palack\'{y} University, 17. listopadu 12,
77146 Olomouc, Czech Republic}

\begin{abstract}
We propose an entanglement distillation and purification scheme for symmetric two-mode entangled Gaussian states
that allows to asymptotically extract a pure entangled Gaussian state from any input entangled symmetric Gaussian state.
The proposed scheme is a modified and extended version of the  entanglement distillation protocol originally developed by [Browne \emph{et al.}, Phys. Rev. A \textbf{67}, 062320 (2003)]. A key feature of the present protocol is that it utilizes a two-copy de-Gaussification
procedure that involves a Mach-Zehnder interferometer with single-mode non-Gaussian filters inserted in its two arms.
The required non-Gaussian filtering operations can be implemented by coherently combining two sequences of 
single-photon addition and subtraction operations.

\end{abstract}


\maketitle

\section{Introduction}

Quantum communication protocols are very sensitive to noise and decoherence that accompanies distribution of quantum states over any realistic quantum channel. The main source of errors in optical quantum communication systems is losses, that e.g. limit the range of current point-to-point quantum cryptography to about one hundred of kilometers \cite{Scarani09}. Significant attention has been therefore paid in recent years to finding schemes that would allow to suppress the noise and decoherence in quantum communication. Of particular interest is a faithful distribution of pure entangled quantum states between two distant parties Alice and Bob who can subsequently use them as resource for some quantum information processing tasks such as quantum teleportation or entanglement-based quantum key distribution. Alice and Bob can achieve this goal by employing a protocol known as entanglement distillation \cite{Bennett96,Deutsch96}. They first distribute several copies of an entangled state over the noisy channel. From the shared copies of a weakly entangled and mixed state they then extract a highly entangled pure state by means of local quantum operations and classical communication (LOCC).  Elementary entanglement distillation and concentration schemes have been successfully demonstrated experimentally  for entangled two-photon states \cite{Kwiat01,Pan03,Zhao03} as well as for two entangled optical modes \cite{Dong08,Hage08,Takahashi10}.

Particularly interesting but also complicated is the distillation of so-called continuous-variable (CV) entanglement where we deal with quantum states belonging to an infinite dimensional Hilbert space of field modes. The experimentally easily accessible CV Gaussian states, which possess Gaussian Wigner function, cannot be distilled by Gaussian operations only \cite{Eisert02,Giedke02,Fiurasek02}. This means that some non-Gaussian operation, such as photon-counting \cite{Opatrny00} or Kerr nonlinearity 
\cite{Duan00,Fiurasek03} is required. An iterative CV entanglement distillation protocol for Gaussian states was proposed by Browne and coworkers \cite{Browne03,Eisert04}.
Their approach combines two steps, see Fig.~1. First, Gaussian states are de-Gaussified by an appropriate quantum filter. Then, the states are re-Gaussified by an iterative procedure whose each steps involves interference of two copies of a two-mode state on balanced beam splitters, followed by projection of one output on a vacuum, c.f. Fig. 1(b).
It can be rigorously proved that this protocol converges to a Gaussian state. This state, however, will generally be mixed, only under very special conditions the protocol will converge to a pure state. The distillation procedure thus increases entanglement of the state, but at the same time it reduces purity of the state, as we shall show below. A possible solution of this problem would be to send through the lossy channel a very weakly entangled pure state whose entanglement 
could then be increased by distillation \cite{Lund09} or probabilistic noiseless amplification \cite{Ralph08,Xiang10,Marek10,Fiurasek09} while preserving high purity of the state. However, this approach is not applicable in situations where
two distant parties share some fixed noisy entangled states and want to extract pure entangled state without having control over the source of the states.

\begin{figure}[!b!]
\centerline{\includegraphics[width=0.7\linewidth]{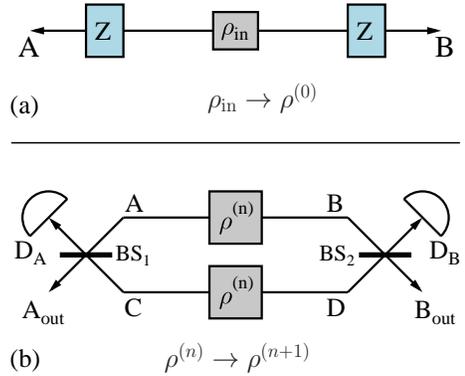}}
\caption{Entanglement distillation scheme for two-mode Gaussian states \cite{Browne03,Eisert04}. (a) De-Gaussification of initial two-mode 
Gaussian state $\hat{\rho}_{\mathrm{in}}$ by local non-Gaussian operations (quantum filters) $\hat{Z}$. (b) Iterative Gaussification procedure whose single step involves local interference of two copies of a state on balanced beam splitters  BS$_1$ and BS$_2$ followed by projections of one output mode on each side onto vacuum using photodetectors  D$_A$ and D$_{B}$, respectively. In the protocol, Alice and Bob use classical communication to announce success or failure  of local filtering operations and projections. }
\end{figure}

In this paper we propose a modified and extended version of  entanglement distillation protocol for symmetric two-mode 
Gaussian states, which asymptotically converges to a pure Gaussian state. This is a crucial advantage with respect to previously proposed protocols for CV entanglement distillation \cite{Browne03,Eisert04,Lund09} and  experimentally demonstrated CV entanglement concentration schemes \cite{Dong08,Takahashi10} that cannot fully suppress the effect of losses and cannot be used to distill \emph{pure} entangled Gaussian state from noisy weakly entangled Gaussian states.
The CV entanglement distillation scheme considered in the present paper has a nested structure. The whole distillation procedure including non-Gaussian  filtration and Gaussification is performed several times, so that already distilled Gaussian states are taken as an input of the next stage of the protocol. Moreover, we significantly alter the non-Gaussian part of the protocol. The de-Gaussification is not performed on a single copy, as in previous works \cite{Opatrny00,Eisert04,Lund09,Takahashi10}, but two copies of the state are used simultaneously and one of them is consumed by the protocol. This makes our approach slightly more resource demanding but also enables us to distill Gaussian states with any chosen amount of entanglement and an arbitrary high purity.

\section{Distillation of Gaussian entanglement}

Gaussian states can be succinctly characterized by first and second moments of quadrature operators. Let $\hat{x}_j$ and $\hat{p}_j$ denote amplitude and phase quadrature operators of mode $j$, respectively. The quadrature operators satisfy canonical commutation relations $[\hat{x}_j,\hat{p}_k]=i\delta_{jk}$. We collect all operators into a vector $\bm{\hat{r}}=(\hat{x}_1,\hat{p}_1,\hat{x}_2,\hat{p}_2,\cdots,\hat{x}_N,\hat{p}_N)$, where $N$ is the total number of modes.  The second moments  can be collected into a  a covariance matrix $\gamma$ whose elements are defined as 
$\gamma_{jk}=\langle \{\Delta\hat{r}_j,\Delta\hat{r}_k \}\rangle $, where $\Delta \hat{r}_j=\hat{r}_j-\langle \hat{r}_j\rangle$ and $\{,\}$ denotes the anticommutator. The first moments $\langle \hat{r}_j\rangle$ can be set to zero by local coherent displacements which do not modify entanglement of the state. In the rest of the paper we will therefore consider only Gaussian states with $\langle \bm{\hat{r}}\rangle =0$.

Let us briefly review the iterative entanglement distillation protocol for Gaussian states proposed by Browne \emph{et al.} \cite{Browne03,Eisert04} and depicted in Fig. 1. In what follows we will focus on a symmetric protocol. More specifically, we will assume that a covariance matrix of a two-mode Gaussian state $\hat{\rho}$ shared initially by Alice and Bob can be expressed as
\begin{equation}
\gamma= 
\left(
\begin{array}{cccc}
C & 0 & S & 0 \\
0 & C & 0 & -S \\
S & 0 & C & 0 \\
0 & -S & 0 & C 
\end{array}
\right).
\label{gamma}
\end{equation}
The generalized Heisenberg uncertainty relation implies that $C \geq 1$ and $C^2-S^2\geq 1$ must hold. Furthermore, we require that the state is entangled, otherwise any distillation would be pointless. This yields another constraint \cite{Lund09},
$C-|S| <1$.
Pure two-mode squeezed vacuum state is recovered for $C=\cosh(2r)$ and $S=\sinh(2r)$, where $r$ is a two-mode squeezing constant. 

As shown by Lund and Ralph \cite{Lund09}, the entangled state (\ref{gamma}) can be conveniently represented as a pure two-mode squeezed vacuum state with some squeezing parameter $r$ transmitted over a lossy channel with intensity transmittance $T$, 
\begin{equation}
\gamma=T \gamma_{\mathrm{TMSV}}+(1-T)\gamma_{\mathrm{vac}},
\end{equation}
where $\gamma_{\mathrm{TMSV}}$ is a covariance matrix of two-mode squeezed vacuum and the covariance matrix of vacuum state $\gamma_{\mathrm{vac}}$ is an identity matrix.
We immediately obtain expressions for $C$ and $S$ in terms of $r$ and $T$,
\begin{equation}
C=T\cosh(2r)+1-T, \qquad S=T\sinh(2r),
\label{CSformula}
\end{equation}
whose inversion gives \cite{Lund09}
\begin{equation}
\tanh(2r)=\frac{2S(C-1)}{S^2+(C-1)^2}, \qquad T=\frac{S^2-(C-1)^2}{2(C-1)}.
\label{rTformula} 
\end{equation}
The symmetric entanglement distillation protocol depicted in Fig. 1 preserves the structure of two-mode state (\ref{gamma}) while altering the parameters $C$ and $S$, or,
equivalently the two-mode squeezing $r$ and the channel transmittance $T$. As shown in Fig.~1(a), a first step of the protocol is probabilistic de-Gaussification of the two-mode state
by  local single-mode filtering operations $\hat{Z}=\hat{n}+w$, where $\hat{n}$ is a photon number operator and $w$ is some constant. Such operation can be implemented by mixing an ancilla 
single photon state $|1\rangle$ with a signal beam on an unbalanced beam splitter with carefully chosen transmittance, 
 followed by projection of ancilla output port onto a single-photon state \cite{Clausen03,Browne03}.
 Note that we can write
\begin{equation}
\hat{Z}= (1-w)\hat{a}^\dagger \hat{a} +w \hat{a} \hat{a}^\dagger,
\end{equation}
where $\hat{a}$ and $\hat{a}^\dagger$ denote annihilation and creation operators. This suggests that,
alternatively, one can coherently combine two sequences of the elementary operations of single-photon addition \cite{Zavatta04} and single-photon subtraction \cite{Ourjoumtsev06,Nielsen06,Wakui07} to implement the desired operation $\hat{Z}$ \cite{Zavatta09,Fiurasek09}. A feasibility of this latter approach has been demonstrated in a recent experiment where a coherent combination of operations $\hat{a}\hat{a}^\dagger$
and $\hat{a}^\dagger\hat{a}$ has been implemented in order to directly  test the commutation relations for bosonic creation and annihilation operators \cite{Zavatta09}.

After local filtering, the resulting states are then re-Gaussified by an iterative protocol. An elementary step of this Gaussification procedure
 consists in local interference of corresponding modes of two copies of the state on balanced beam splitters BS$_1$ and BS$_2$ followed by projection of one output port of each beam splitter onto vacuum state, see Fig. 1(b). Mathematically, the Gaussification protocol is described by an iterative map \cite{Browne03,Eisert04},
\begin{equation}
\hat{\rho}^{(n+1)}=\frac{\mathcal{E}_G\left(\hat{\rho}^{(n)}\otimes \hat{\rho}^{(n)}\right) }{\mathrm{Tr}\left[\mathcal{E}_G(\hat{\rho}^{(n)}\otimes \hat{\rho}^{(n)})\right]},
\end{equation}
where $\mathcal{E}_G$ denotes a probabilistic Gaussian operation on two copies of a two-mode state $\hat{\rho}^{(n)}$. 
Using the labeling of modes as in Fig. 1(b) we can write
\begin{equation}
\mathcal{E}_G(\hat{\rho}_{ABCD})=  \hat{K}^\dagger \hat{\rho}_{ABCD} \hat{K},
\end{equation}
where $\hat{K}= \hat{U}_{\mathrm{BS},AC}^\dagger\otimes \hat{U}_{\mathrm{BS},BD}^\dagger |0,0\rangle_{CD}$, $|0\rangle$ denotes vacuum state, and $\hat{U}_{\mathrm{BS}}$ stands for a two-mode unitary operation corresponding to interference of two modes on a balanced beam splitter. If the Gaussification procedure converges, then the asymptotic state $\hat{\rho}^{\infty}$ is Gaussian, and its covariance matrix can be determined from a state $\hat{\rho}^{(1)}$ after the first iteration of $\mathcal{E}_{G}$.  We use short hand notation $\rho_{jk,mn}=\langle j,k|\hat{\rho}|m,n\rangle$ for density matrix elements in Fock basis. We define auxiliary matrix 
\begin{equation}
\hat{\sigma}=\frac{\hat{\rho}^{(1)}}{\rho_{00,00}^{(1)}}.
\end{equation}
Covariance matrix $\gamma_{\mathrm{out}}$ of asymptotic Gaussian state $\hat{\rho}^{\infty}$ is then a function of the matrix elements $\sigma_{jk,mn}$ with $j+k+m+n=2$. Explicit expressions and details can be found in Refs. \cite{Eisert04,Lund09}. For the class of symmetric states (\ref{gamma}) considered here some of the matrix elements are equal to zero, 
\begin{equation}
\sigma_{20,00}=\sigma_{02,00}=\sigma_{00,20}=\sigma_{00,02}=\sigma_{10,01}=\sigma_{01,10}=0,
\end{equation}
and furthermore $\sigma_{10,10}=\sigma_{01,01}$. As we shall see this property is satisfied also by our generalized distillation protocol that will be described in Seciton IV.

A crucial observation made by Lund and Ralph \cite{Lund09} is that the distillation protocol shown in Fig. 1 does not change the value of a parameter
\begin{equation}
\epsilon=(1-T)\tanh(r) \equiv \frac{C^2-S^2-1}{2S}.
\label{epsdefinition}
\end{equation}
It holds that $\epsilon$ is the same for an initial noisy Gaussian state before distillation $\hat{\rho}_{\mathrm{in}}$ and the final distilled Gaussian state $\hat{\rho}^{\infty}$. The distillation simultaneously increases both effective two-mode squeezing parameter $r$ and effective transmittance $T$ while keeping $\epsilon$ constant. This clearly increases the entanglement of the state. However, a simple calculation shows that a purity of the state defined as $\mathcal{P}=\mathrm{Tr}(\hat{\rho}^2)$ is reduced by the distillation. For Gaussian states it holds that $\mathcal{P}=1/\sqrt{\det(\gamma)}$. With the help of formulas (\ref{gamma}), (\ref{CSformula}) and (\ref{epsdefinition}) we obtain 
\begin{equation}
\mathcal{P}=\left[1-2\epsilon^2-2\epsilon^2\cosh(2r)+2\epsilon \sinh(2r)\right]^{-1}.
\end{equation}
Since $\epsilon <1$ by definition, $\mathcal{P}$ is a decreasing function of $r$ for a fixed $\epsilon$. This shows that the entanglement distillation protocol decreases state purity. For symmetric Gaussian states (\ref{gamma}) the von Neumann entropy of the state depends only on $\mathcal{P}$ and is a monotonically decreasing function of $\mathcal{P}$. So the distillation also increases von Neumann entropy of the two-mode Gaussian state. 

\section{Single-copy pre-processing}

In order to increase the purity of distilled state while preserving (or even increasing) its entanglement, we need to find a way how to reduce the value of parameter $\epsilon$. We seek to achieve this by a suitable modification of the de-Gaussification step of the distillation protocol. We will first consider two simple single-copy strategies. One option is to replace filter $\hat{Z}=\hat{n}+w$ with a single-photon subtraction $\hat{Z}=\hat{a}$ \cite{Opatrny00}. Another option is  to perform some local Gaussian operations prior to de-Gaussification. However, as we shall argue below, these approaches do not allow to decrease value of $\epsilon$. In Section IV. we therefore develop an alternative collective two-copy de-Gaussification scheme that does the job and reduces value of $\epsilon$ of the final distilled Gaussian state.

\subsection{Local single-photon subtraction}

If we use single-photon subtraction as the de-Gaussification filter $\hat{Z}$, then the initial two-mode density matrix $\hat{\rho}_{\mathrm{in}}$ will be transformed as,
\begin{equation}
\hat{\rho}^{(0)}= \frac{\hat{a}\hat{b}\,\hat{\rho}_{\mathrm{in}}\,\hat{a}^\dagger \hat{b}^\dagger}{\mathrm{Tr}[\hat{a}^\dagger\hat{a}\,\hat{b}^\dagger\hat{b}\,\hat{\rho}_{\mathrm{in}}]},
\end{equation}
where $\hat{a}$ and $\hat{b}$ denote annihilation operator of Alice's and Bob's mode of shared two-mode state, respectively.
For symmetric mixed Gaussian state with covariance matrix (\ref{gamma}) the matrix elements $\sigma_{jk,mn}$ required for determination of final distilled Gaussian state can be analytically calculated. The nonzero elements read
\begin{eqnarray}
&&\sigma_{11,00}=\sigma_{00,11}=\frac{2T\lambda\left[1+2\lambda^2(1-T)^2\right]}{1-\lambda^4(1-T)^4}, \nonumber \\
&&\sigma_{10,10}=\sigma_{01,01}=\frac{2T(1-T)\lambda^2\left[2+\lambda^2(1-T)^2\right]}{1-\lambda^4(1-T)^4}, \nonumber \\
\label{sigmaSPS}
\end{eqnarray}
where $\lambda=\tanh r$. With these expressions at hand we can analytically calculate covariance matrix $\gamma_{\mathrm{out}}$ of the distilled Gaussian state $\hat{\rho}^{(\infty)}$. 
The whole protocol preserves the symmetry of the state and $\gamma_{\mathrm{out}}$ has the form (\ref{gamma}), 
only the value of parameters $C$ and $S$ is changed. We are primarily interested in $\epsilon_{\mathrm{out}}$. After some algebra we find that this parameter can be expressed as
a ratio of two elements of matrix $\sigma$,
\begin{equation}
\epsilon_{\mathrm{out}}=\frac{\sigma_{10,10}}{\sigma_{11,00}}.
\label{epsilon_sigma}
\end{equation}
On inserting the explicit formulas (\ref{sigmaSPS}) into Eq. (\ref{epsilon_sigma}) we obtain
\begin{equation}
\frac{\epsilon_{\mathrm{out}}}{\epsilon_{\mathrm{in}}}=\frac{2+\lambda^2(1-T)^2}{1+2\lambda^2(1-T)^2},
\label{epsilonSPS}
\end{equation}
where $\epsilon_{\mathrm{in}}=(1-T)\lambda$.
Since $\lambda^2(1-T)^2 <1 $ we have $\epsilon_{\mathrm{out}} \geq \epsilon_{\mathrm{in}}$. The replacement of Fock-diagonal filter $\hat{n}+w$ with photon subtraction operation $\hat{a}$ makes things even worse in the sense that the value of $\epsilon$ is increased instead of decreased.

\subsection{Local Gaussian operations}

Experimentally simplest filtering would be based on application of local Gaussian operations and measurements.
One may conjecture that local Gaussian operations might help to reduce value of $\epsilon$ although they cannot increase entanglement of the state. Let us therefore investigate this possibility. Suppose that we perform a local Gaussian operation on each mode of the state $\hat{\rho}_{\mathrm{in}}$ prior to the de-Gaussification by filter $\hat{Z}$. A general two-mode Gaussian operation can be characterized by a four-mode covariance matrix $\Gamma$ that can be conveniently split into output  and input parts labeled by indices 1 and 2, respectively,
\begin{equation}
\Gamma= \left( 
\begin{array}{cc}
\Gamma_1 &  \Gamma_{12} \\
\Gamma_{12}^T & \Gamma_{2} 
\end{array}
\right).
\end{equation}
This operation transforms covariance matrix of a Gaussian state according to the formula \cite{Giedke02}
\begin{equation}
\gamma'= \Gamma_{1}-\Gamma_{12}\left[\Gamma_2+\Sigma \gamma_{AB}\Sigma^T \right]^{-1} \Gamma_{12}^T,
\label{GaussianCPmap}
\end{equation}
where $\Sigma=\mathrm{diag}(1,-1,1,-1)$.
For local operations the matrix $\Gamma$ reduces to a direct sum of two $4 \times 4$ two-mode covariance matrices,
\begin{equation}
\Gamma=\Gamma_{AA'}\oplus \Gamma_{BB'},
\end{equation}
where the unprimed (primed) labels refer to input (output) modes. We require that the transformation (\ref{GaussianCPmap}) preserves the form and symmetry of the covariance matrix (\ref{gamma}). This is guaranteed when  $\Gamma_{AA'}=\Gamma_{BB'}$ and both these matrices have the same structure as $\gamma$ in Eq. (\ref{gamma}). Moreover it suffices to consider only covariance matrices corresponding to pure Gaussian states because any operation represented by a covariance matrix $\Gamma$ of a mixed state can be obtained as a Gaussian mixture of operations represented by pure-state covariance matrices $\Gamma$. This implies that $\Gamma_{AA'}=\Gamma_{BB'}$ are covariance matrices 
of a pure two-mode squeezed vacuum state with some squeezing constant $s$.
Physically, the resulting operation (\ref{GaussianCPmap}) can be implemented by mixing each mode with a vacuum state on an unbalanced beam splitter with a properly chosen amplitude transmittance $\tau$ and projecting one output mode on a vacuum state. The parameters $C'$ and $S'$ of covariance matrix of a symmetric Gaussian state after application of the local Gaussian operations can be expressed as follows, 
\begin{eqnarray*}
C'&=&\frac{C[\cosh^2(2s)+1]+(C^2-S^2+1)\cosh(2s)}{(C+\cosh(2s))^2-S^2}, \\
S'&=&\frac{S\sinh^2(2s)}{(C+\cosh(2s))^2-S^2}. 
\end{eqnarray*}
On inserting these expressions into Eq. (\ref{epsdefinition}) we find after lengthy but straightforward calculation that $\epsilon'=\epsilon$, hence this Gaussian operation does
not change value of $\epsilon$. Another way of proving this statement is based on the observation that the considered local Gaussian operation is a filter diagonal in Fock basis,
$\hat{Z}_G=\tau^{\hat{n}}$. For a symmetric Gaussian state (\ref{gamma}) the parameter $\epsilon$ can be expressed as 
\begin{equation}
\epsilon=\frac{\rho_{10,10}}{\rho_{11,00}},
\label{epsilonrho}
\end{equation}
 c.f. Eq. (\ref{epsilon_sigma}). 
Density matrix elements of un-normalized state after local Gaussian filters $\hat{Z}_G$ are given by ${\rho}_{10,10}'=\tau^2{\rho}_{10,10}$, ${\rho}_{11,00}'=\tau^2{\rho}_{11,00}$, and it immediately 
follows that the ratio ${\rho}_{10,10}'/{\rho}_{11,00}'={\rho}_{10,10}/{\rho}_{11,00}$ remains unchanged. 
Note that the above analysis was based on certain symmetry assumptions so it does not exhaustively cover all possible local Gaussian operations. Still, it provides a strong indication that local Gaussian operations cannot help to reduce value of $\epsilon$.

\section{Two-copy de-Gaussification}

\begin{figure}[t]
\includegraphics[width=0.95\linewidth]{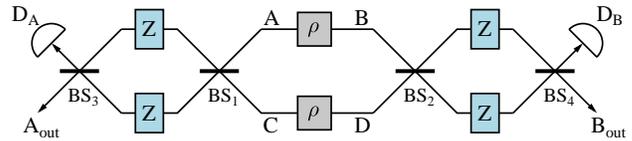}
\caption{Two-copy de-Gaussification scheme. BS$_j$ denotes  balanced beam splitter, Z indicates a 
non-Gaussian operation -- a Fock state filter $\hat{Z}=\hat{n}-1$, and each detection block D$_j$ makes projection onto 
a superposition of vacuum and single-photon state.}
\end{figure}

We now present  a  two-copy de-Gaussification scheme that reduces the factor $\epsilon$, see Fig. 2. 
The initial state shared by Alice and Bob can be written as $\hat{\rho}_{\mathrm{in},AB}\otimes \hat{\rho}_{\mathrm{in},CD}$ with modes A and C belonging to Alice and modes B and D belonging to Bob.
Both Alice and Bob send their two modes through a Mach-Zehnder interferometer formed by two balanced beam splitters BS$_1$ and BS$_3$ (BS$_2$ and BS$_4$). A non-Gaussian operation is applied to each mode inside the interferometer.
 We choose a Fock-state filter $\hat{Z}_1=\hat{n}-1$ that eliminates the single-photon state, $\hat{Z}_1|1\rangle=0$. 
Finally, the output modes C and D are projected onto a superposition of vacuum and single-photon state 
\begin{equation}
|q\rangle= \frac{1}{\sqrt{1+q^2}}(q|0\rangle+|1\rangle),
\label{qstate}
\end{equation}
where we assume real $q$. By changing $q$ we can control the amount of entanglement of the final distilled Gaussian state.
Projection onto state (\ref{qstate}) can be accomplished by a sequence of a coherent displacement $\hat{D}(q)$, single-photon subtraction, another coherent displacement $\hat{D}^\dagger(q)$ and projection onto vacuum \cite{Dakna99}. In mathematical terms, we have 
\begin{equation}
\langle q|\propto \langle 0| \hat{D}^\dagger(q) \hat{a} \hat{D}(q)= \langle 0| (\hat{a}+q).
\end{equation}

To grasp the principle of the scheme shown in Fig. 2 let us first consider distillation of truncated two-mode squeezed states
\begin{equation}
|\Lambda\rangle=\frac{1}{\sqrt{1+\lambda^2}}(|00\rangle+\lambda|11\rangle)
\label{truncated}
\end{equation}
transmitted from the central source to Alice and Bob over two identical lossy channels with intensity transmittance $T$. 
In the basis $|00\rangle$, $|10\rangle$, $|01\rangle$, $|11\rangle$, the density matrix of  state shared by Alice and Bob reads
\begin{equation}
\hat{\rho}_{\mathrm{in}}=\frac{1}{1+\lambda^2}\left(
\begin{array}{cccc}
1+\lambda^2 R^2 & 0 & 0 & \lambda T \\
0 & \lambda^2 TR & 0 & 0 \\
0 & 0 & \lambda^2 TR & 0  \\
\lambda T & 0 & 0 & \lambda^2 T^2
\end{array}
\right),
\label{rhotruncated}
\end{equation}
where $R=1-T$.
Parameter $\epsilon$ of a Gaussian state that would be obtained from this resource by a standard Gaussification procedure as shown in Fig. 1 
can be directly calculated using Eq. (\ref{epsilonrho}) and we obtain $\epsilon=(1-T)\lambda$, as expected. Suppose now that two copies of the state (\ref{rhotruncated}) are used as an input of the
de-Gaussification scheme of Fig. 2. It turns out that the interference on balanced beam splitters followed by filters $\hat{n}-1$ suppresses all contributions stemming from the loss of an odd number of photons (one or three) during the transmission. After some algebra we obtain density matrix of the state after de-Gaussification,
\begin{widetext}
\begin{equation}
\hat{\rho}_{AB}'=\frac{1}{q^4(1+2\lambda^2R^2)+\lambda^4(q^2R^2+T^2)^2}\left(
\begin{array}{cccc}
q^4(1+\lambda^2R^2)^2 & 0 & 0 & \lambda^2 T^2 q^2 \\
0 & \lambda^4 T^2R^2 q^2 & 0 & 0 \\
0 & 0 & \lambda^4 T^2R^2 q^2 & 0  \\
\lambda^2 T^2 q^2 & 0 & 0 & \lambda^4 T^4
\end{array}
\right).
\end{equation}
\label{rhodegaussified}
\end{widetext}
We can determine the parameter $\epsilon'$ of this state similarly as before for $\hat{\rho}_{\mathrm{in}}$ and we find
\begin{equation}
\epsilon'= (1-T)^2\lambda^2.
\label{epsilonprime}
\end{equation}
This means that $\epsilon'=\epsilon^2$ and since $\epsilon<1$ we have $\epsilon'<\epsilon$ as desired.

We next prove that this property remains true even for input symmetric Gaussian states (\ref{gamma}). The overall transformation induced by the filtration can be expressed as
\begin{equation}
\hat{\rho}_{AB}'=(\hat{F}_{AC}^\dagger \otimes \hat{F}_{BD}^\dagger)  \hat{\rho}_{\mathrm{in},AB}\otimes \hat{\rho}_{\mathrm{in},CD} (\hat{F}_{AC} \otimes \hat{F}_{BD}),
\end{equation}
where the local two-mode filters are identical,   $\hat{F}_{AC}=\hat{F}_{BD}$,
 and e.g. the filter on Alice's side can be expressed as
\begin{equation}
\hat{F}_{AC}=\hat{U}_{\mathrm{BS},AC}^\dagger(\hat{n}_A-1)(\hat{n}_C-1)\hat{U}_{\mathrm{BS},AC}|q\rangle_C.
\end{equation}
The balanced beam splitter transforms input annihilation operators into their balanced combinations, 
\begin{eqnarray}
\hat{U}_{\mathrm{BS},AC}^\dagger \hat{a} \hat{U}_{\mathrm{BS},AC} &=& \frac{1}{\sqrt{2}}(\hat{a}+\hat{c}), \nonumber \\
\hat{U}_{\mathrm{BS},AC}^\dagger \hat{c} \hat{U}_{\mathrm{BS},AC} &=& \frac{1}{\sqrt{2}}(\hat{a}-\hat{c}), 
\label{UBSac}
\end{eqnarray}
With the help of transformation rules (\ref{UBSac}) we obtain after some algebra
\begin{eqnarray}
\hat{F}_{AC}&=&
\frac{q}{4}(\hat{n}_A-1)( \hat{n}_A-4)|0\rangle_C
+ \frac{1}{4}\hat{n}_A(\hat{n}_A-5)|1\rangle_C \nonumber \\
& & - \frac{1}{4}\hat{a}^2 \left(q\sqrt{2}|2\rangle_C+\sqrt{6}|3\rangle_C\right).
\end{eqnarray}
The filtration effectively replaces projection onto a Fock state $|n\rangle_A$ with projection onto state $\hat{F}|n\rangle_A$.
In particular, for the three lowest Fock states we obtain the mapping
\begin{eqnarray*}
|0\rangle_A & \rightarrow & q|0,0\rangle_{AC}, \\
|1\rangle_A & \rightarrow & -|1,1\rangle_{AC}, \\ 
|2\rangle_A & \rightarrow & -\frac{1}{2}|2\rangle_A(q|0\rangle_{C}+3|1\rangle_{C}) \\
& & -\frac{1}{2}|0\rangle_{A} (q|2\rangle_{C}+\sqrt{3}|3\rangle_{C}).
\end{eqnarray*}
Similar formulas hold for filter on Bob's side. Due to the symmetric structure of the Gaussian state (\ref{gamma}), the only nonzero density matrix elements
of state $\hat{\rho}'$ relevant for determination of the asymptotic Gaussian state after Gaussification procedure are the following, 
\begin{eqnarray}
& {\rho}_{00,00}'=q^4{\rho}_{00,00}^2, & \nonumber \\ 
& {\rho}_{10,10}'={\rho}_{01,01}'=q^2{\rho}_{10,10}^2, & \nonumber  \\
& {\rho}_{11,00}'={\rho}_{00,11}'=q^2{\rho}_{11,00}^2. &
\label{rhosquared}
\end{eqnarray}
If we insert matrix elements (\ref{rhosquared}) into Eq. (\ref{epsilonrho}) we find that indeed the parameter $\epsilon_{\mathrm{out}}$ of a Gaussian state obtained by Gaussification of 
$\hat{\rho}'$ is equal to the square of $\epsilon_{\mathrm{in}}$ of the initial Gaussian state $\hat{\rho}_{\mathrm{in}}$,
\begin{equation}
\epsilon_{\mathrm{out}}=\epsilon_{\mathrm{in}}^2.
\end{equation}

\begin{figure}[t]
\includegraphics[width=0.75\linewidth]{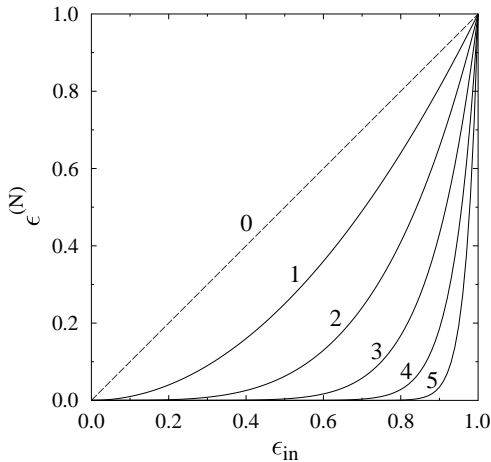}
\caption{Parameter $\epsilon^{(N)}$ of distilled Gaussian state after $N$ stages of nested entanglement 
distillation protocol is plotted as a function of $\epsilon_{\mathrm{in}}$. The numerical labels indicate 
number of stages $N$ of the protocol. }
\end{figure}

By iterating the whole entanglement distillation protocol several times the 
parameter $\epsilon$ can be reduced to an arbitrarily small value.
Such a nested distillation protocol is similar to a quantum repeater scheme \cite{Briegel98,Duan01}. Alice and Bob divide the shared copies of states $\hat{\rho}_{\mathrm{in}}$ into many blocks.
They perform distillation on each block, thereby obtaining several copies of distilled state $\hat{\rho}^{\infty}$. These states are then used as an input for the next stage 
of the nested distillation scheme. After $N$ stages of this protocol the parameter $\epsilon$ is reduced to
\begin{equation}
\epsilon^{(N)}=\epsilon_{\mathrm{in}}^{2^N}.
\end{equation}
This dependence is plotted in Fig. 3. We can see that the protocol quickly  converges to $\epsilon=0$ corresponding to effective channel transmittance $T_{\mathrm{eff}}=1$ and a pure state. 

Symmetric two-mode Gaussian states can be characterized by their purity $\mathcal{P}$ and entanglement of formation 
$E_f$ that can be calculated analytically for this class of states \cite{Giedke09}.
In Fig. 4 we plot purity and entanglement of formation of Gaussian state obtained after $N$ stages of the nested distillation protocol as a function of the initial effective channel transmittance $T$. At each stage of the protocol, $q$ is adjusted such that the effective two-mode squeezing given by Eq. (\ref{rTformula}) remains constant, $r=1$. Each stage of the protocol thus increases the entanglement of the state, see Fig. 4(b). Purity of the state is also increased, except for the region of very low $T$, where state purity can be actually reduced after the first stage of the protocol. Nevertheless, after subsequent stages purity increases and approaches asymptotic value $\mathcal{P}=1$.

\begin{figure}[b]
\includegraphics[width=0.8\linewidth]{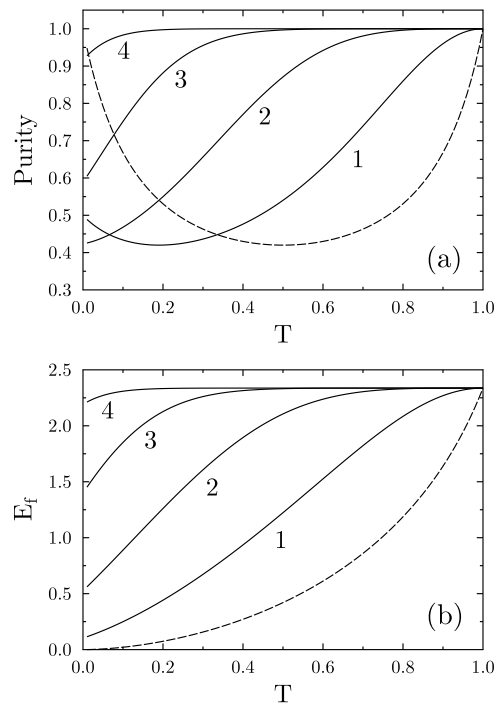}
\caption{Purity (a) and entanglement of formation (b) of distilled Gaussian state after $N$ stages of nested entanglement 
distillation protocol is plotted as a function of the effective channel transmittance $T$ for a fixed effective two-mode squeezing 
constant $r=1$. The numerical labels indicate number of stages $N$ of the protocol, dashed lines represent values of $\mathcal{P}$ and $E_f$
for the initial state before distillation.}
\end{figure}

\section{Conclusions}

In this paper we have proposed an entanglement distillation scheme for symmetric Gaussian states that allows to asymptotically extract a pure entangled Gaussian 
state from an arbitrary input entangled symmetric Gaussian state. The present scheme is a modified and extended version of the entanglement distillation protocol 
for Gaussian states developed by Browne and coworkers \cite{Browne03,Eisert04}. We employ a two-copy de-Gaussification procedure and a nested entanglement distillation scheme where outputs 
of a full iterative distillation protocol are used as inputs of the next stage of the nested scheme.  To make the presentation as transparent and comprehensible 
as possible we have considered an idealized scenario where all local operations and measurements are implemented perfectly, i.e. without any errors. 
In practice, both photon addition and subtraction as well as interference on beam splitters and projection onto vacuum state will suffer from some errors. 
Also, any practical implementation of Gaussification will be only approximate, with finite number of iterations. The number of iterations will influence both
the convergence and success rate of the protocol.
An analysis of the influence of all these effects on the performance of the protocol will be a subject of the future work.

The entanglement distillation protocol discused in this paper combines several elementary building blocks, each of which has been already successfully demonstrated experimentally. The Gaussification of two and three copies of two-mode entangled states of light by their local interference on beam splitters and Gaussian measurements on some output ports has been employed to experimentally distill entanglement of phase-diffused two-mode squeezed states \cite{Hage08,Hage10}. The required non-Gaussian operation $\hat{Z}=\hat{n}-1=2\hat{a}^\dagger\hat{a}-\hat{a}\hat{a}^\dagger$ is a linear combination of two sequences of photon addition and photon subtraction that has been recently implemented experimentally for the purpose of direct experimental testing of fundamental quantum commutation relations \cite{Zavatta09}. In view of these recent achievements, a proof-of-principle demonstration of the entanglement distillation scheme proposed in this paper appears to be within the current experimental limits. However, the purely optical scheme is likely to exhibit 
exponential decrease of success probability with growing number of distilled copies. This unfavourable scaling could be avoided with the help of a quantum memory \cite{Lvovsky09,Hammerer10}, similarly to the quantum repeater schemes \cite{Briegel98,Duan01}.

\acknowledgments

This work was supported by MSMT under projects LC06007, MSM6198959213, and 7E08028, by
the EU under the FET-Open project COMPAS (212008), and also by GACR under project GA202/08/0224.


\begin{thebibliography}{99}

\bibitem{Scarani09}
V. Scarani, H. Bechmann-Pasquinucci, N.J. Cerf, M. Du\v{s}ek, N. L\"{u}tkenhaus, and M. Peev,
Rev. Mod. Phys. \textbf{81}, 1301 (2009). 

\bibitem{Bennett96}
C.H. Bennett, G. Brassard, S. Popescu, B. Schumacher, J.A. Smolin, and W.K. Wootters,  Phys. Rev. Lett. \textbf{76}, 722 (1996).

\bibitem{Deutsch96}
D. Deutsch et al., Phys. Rev. Lett. \textbf{77}, 2818 (1996).

 
\bibitem{Kwiat01}
P.G. Kwiat, S. Barraza-Lopez, A. Stefanov, and N. Gisin,  Nature \textbf{409}, 1014-1017 (2001).

\bibitem{Pan03}
J.W. Pan, S. Gasparoni, R. Ursin, G. Weihs, and A. Zeilinger, Nature \textbf{423}, 417 (2003).


\bibitem{Zhao03}
Z. Zhao, T. Yang, Y.A. Chen, A.N. Zhang, and J.W. Pan, Phys. Rev. Lett. \textbf{90}, 207901 (2003).


\bibitem{Hage08} 
B. Hage, A. Samblowski, J. DiGuglielmo, A. Franzen, J. Fiur\'{a}\v{s}sek, and R. Schnabel, Nature Physics 4, 915 (2008).

\bibitem{Dong08}
R. Dong, M. Lassen, J. Heersink, C. Marquardt, R. Filip, G. Leuchs, and U.L. Andersen, Nature Phys. \textbf{4}, 919 (2008).


\bibitem{Takahashi10}
H. Takahashi, J.S. Neergaard-Nielsen, M. Takeuchi, M. Takeoka, K. Hayasaka, A. Furusawa, and M. Sasaki,  
Nature Photonics \textbf{4}, 178 (2010).


\bibitem{Eisert02}
J. Eisert, S. Scheel, and M.B. Plenio, Phys. Rev. Lett. \textbf{89}, 137903 (2002).

\bibitem{Giedke02}
G. Giedke and J.I. Cirac, Phys. Rev. A \textbf{66}, 032316 (2002).

\bibitem{Fiurasek02}
J. Fiur\'{a}\v{s}ek, Phys. Rev. Lett. \textbf{89}, 137904 (2002).

\bibitem{Opatrny00}
T. Opatrn\'{y}, G. Kurizki, and D.-G. Welsch, Phys. Rev. A \textbf{61}, 032302 (2000).


\bibitem{Duan00}
L.M. Duan, G. Giedke, J.I. Cirac, and P. Zoller, Phys. Rev. Lett. \textbf{84}, 4002 (2000).

\bibitem{Fiurasek03}
J. Fiur\'{a}\v{s}ek, L. Mi\v{s}ta, Jr., and R. Filip, Phys. Rev. A \textbf{67}, 022304 (2003). 

\bibitem{Browne03}
D.E. Browne, J. Eisert, S. Scheel, and M.B. Plenio, Phys. Rev. A \textbf{67}, 062320 (2003).

\bibitem{Eisert04}
 J. Eisert, D.E. Browne, S. Scheel, and M.B. Plenio, Ann. Phys. \textbf{311}, 431 (2004).

 \bibitem{Lund09}
A.P. Lund and T.C. Ralph, Phys. Rev. A \textbf{80}, 032309 (2009).


\bibitem{Ralph08}
T.C. Ralph and A.P. Lund, in \emph{Quantum Communication Measurement and Computing}, 
Proceedings of 9th International Conference, Ed. A. Lvovsky, 155-160 (AIP, New York 2009); arXiv:0809.0326.  


\bibitem{Xiang10}
G. Y. Xiang, T. C. Ralph, A. P. Lund, N. Walk, and G. J. Pryde, Nature Photonics \textbf{4}, 316 (2010). 

\bibitem{Marek10}
P. Marek and R. Filip, Phys. Rev. A \textbf{81}, 022302 (2010).

\bibitem{Fiurasek09}
J. Fiur\'{a}\v{s}ek, Phys. Rev. A \textbf{80}, 053822 (2009). 


\bibitem{Clausen03}
J. Clausen, L. Knoll, and D.-G. Welsch, Phys. Rev. A \textbf{68}, 043822 (2003).


 \bibitem{Zavatta04}
A. Zavatta, S. Viciani, and M. Bellini, Science \textbf{306}, 660 (2004). 


\bibitem{Ourjoumtsev06}
A. Ourjoumtsev, R. Tualle-Brouri, J. Laurat, and Ph. Grangier, Science \textbf{312}, 83 (2006).


\bibitem{Nielsen06}
J.S. Neergaard-Nielsen, B.M. Nielsen, C. Hettich, K. Molmer, and E.S. Polzik, 
Phys. Rev. Lett. \textbf{97}, 083604 (2006).


\bibitem{Wakui07} 
K. Wakui, H. Takahashi, A. Furusawa, and M. Sasaki, Opt. Express \textbf{15}, 3568 (2007).


\bibitem{Zavatta09}
A. Zavatta, V. Parigi, M. S. Kim, H. Jeong, and M. Bellini, Phys. Rev. Lett. \textbf{103}, 140406 (2009).

\bibitem{Dakna99}
M. Dakna, J. Clausen, L. Kn\"{o}ll, and D.-G. Welsch, Phys. Rev. A \textbf{59}, 1658 (1999); Phys. Rev. A \textbf{60}, 726 (1999).


\bibitem{Briegel98}
H.J. Briegel, W. D\"{u}r, J.I. Cirac, and P. Zoller, Phys. Rev. Lett. \textbf{81}, 5932 (1998).

\bibitem{Duan01}
L.M. Duan, M.D. Lukin, J.I. Cirac, and P. Zoller, Nature \textbf{414}, 413 (2001).

\bibitem{Giedke09}
G. Giedke, M. M. Wolf, O. Kr\"{u}ger, R.F. Werner, and J.I. Cirac, Phys. Rev. Lett. \textbf{91}, 107901 (2003).

\bibitem {Hage10}
B. Hage, A. Samblowski, J. DiGuglielmo, J. Fiur\'{a}\v{s}ek, and R. Schnabel, e-print arXiv:1007.1508; to appear in Phys. Rev. Lett. 

\bibitem{Lvovsky09}
A.I. Lvovsky, B.C. Sanders, and W. Tittel, Nature Phot. \textbf{3}, 706 (2009).

\bibitem{Hammerer10}
K. Hammerer, A.S. S{\o}rensen, and E.S. Polzik, Rev. Mod. Phys. \textbf{82}, 1041 (2010).


\end{thebibliography}
\end{document}